# Study Protocol: Shared Achievements: Exploring the Design of Gameful Collaborative Elements and Fostering Social Relatedness through Team Effort Contributions in a Social Physical Activity App

FAITH YOUNG, Ludwig Boltzmann Institute for Digital Health and Prevention, Austria

DMITRY ALEXANDROVSKY, Karlsruher Institut für Technologie, Germany

DANIELA WURHOFER, Ludwig Boltzmann Institute for Digital Health and Prevention, Austria

EVA-MARIA KRAH, Ludwig Boltzmann Institute for Digital Health and Prevention, Austria

JAN D. SMEDDINCK, Ludwig Boltzmann Institute for Digital Health and Prevention, Austria

This study protocol outlines the design and methodology of a research study investigating collaborative game elements to promote physical activity within digital health interventions. The study aims to examine how social relatedness influences motivation and adherence to step-count goals. Participants will use Shared Achievements, a minimalistic multiplayer step counter game, over two weeks—one week contributing absolute step counts and another sharing step counts as a percentage of a team goal. Data will be collected through usage metrics and participant feedback to evaluate engagement, motivation, and perceived challenges. Findings will inform the design of digital health tools that balance competition and collaboration, optimizing social and behavioural support mechanisms.

## 1 INTRODUCTION

Unhealthy behaviours such as a sedentary lifestyle and inadequate diet are prevalent today and present a societal issue [29]. To prevent avoidable medical conditions, health research suggests that adults should practice regular physical activity (PA) [17, 27, 37]. While many accessible PA and health interventions exist, establishing and pursuing conscious and sustainable behaviour change remains challenging. Internet-based and mobile health interventions (IMIs) grounded in Self-Determination Theory (SDT) [30] and adjacent theories on player experience provide evidence for gameful design patterns as valuable strategies to support motivation to pursue a healthy lifestyle [4, 7, 12]. However, many

Authors' Contact Information: Faith Young, Ludwig Boltzmann Institute for Digital Health and Prevention, Salzburg, Austria, faith.young@lbg.ac.at; Dmitry Alexandrovsky, Karlsruher Institut für Technologie, Karlsruhe, Germany, dmitry.alexandrovsky@kit.edu; Daniela Wurhofer, Ludwig Boltzmann Institute for Digital Health and Prevention, Salzburg, Austria, daniela.wurhofer@lbg.ac.at; Eva-Maria Krah, Ludwig Boltzmann Institute for Digital Health and Prevention, Salzburg, Austria, eva-maria.krah@lbg.ac.at; Jan D. Smeddinck, Ludwig Boltzmann Institute for Digital Health and Prevention, Salzburg, Austria, jan.smeddinck@lbg.ac.at.

so-called "serious games", exergames, or games for change and health interventions [31] suffer from high dropout rates, and gameful design often does not support long-term adherence to the intended program [3]. Collaborative play is known to yield enjoyment and provide meaning to an activity [19, 33], which could justify shifting the focus of gamification research towards gameful design [11] that facilitates collaborative behaviour [18]. Social interaction, self-efficacy, and a sense of relatedness can promote engagement and give meaning to activities [13, 22, 30], suggesting that creating memorable and positive experiences through social collaboration could help people adhere to PA and a healthier lifestyle.

Under the umbrella of health prevention, this work investigates gameful design patterns [24] that build on social interaction and can be employed in digital tools aimed at improving the quality of life by helping people to engage and adhere to walking as a form of PA. This work builds on existing research on fostering PA and is configured to enable the comparative exploration of social relatedness as a main motivational trigger for reaching PA goals. Using a decisively minimalistic multiplayer step counter game to avoid diluting any potential effects gained from the key social aspects, we investigated how collaborative game elements affect the player experience and what social dynamics emerge in a collaborative fitness game. Even in generally collaborative settings, potential sensitivities around more openly sharing PA achievements in a manner that may bring along competitive elements motivated comparison with more indirect forms (here: relative percentage contributions) of sharing and communicating PA achievements as contributions to shared goals [25]. Therefore, this research aims to gauge the general acceptance of the concept of shared achievements to motivate PA as a whole, both in terms of the application design and acceptance as well as the intended increase in steps/PA and/or positive experience linked to PA among participants. The guiding research questions are as follows:

RQ1: How do gameful design patterns, particularly collaborative goals (shared achievements), influence user motivation and adherence to PA in a multiplayer step counter game?

RQ2.1: How do social dynamics within an exergame, particularly exploring differences between absolute (number of steps) and relative (percentage of goal) communication on individual contributions to the team achievements, impact user motivation and adherence to PA goals?

RQ2.2: What role does communication between team members play in fostering engagement in digital health tools?

In line with the study design by Vella et al. [36], we investigate these research aims in an exploratory user study, where the participants use a step counter game for two weeks. By examining the interplay between social relatedness and collaborative game mechanics from a qualitative perspective, this paper contributes to the body of knowledge on how digital health interventions can be more effectively designed to sustain user engagement and motivation. Our findings offer insights into balancing competition and collaboration in gameful design, paving the way for more personalized and effective activity promotion tools.

## 2   RELATED WORK

Digital health interventions are increasingly being used and estimates show that a significant proportion of internet-connected adults in western countries use digital technology to support and monitor their well-being [28].

## 2.1 Playful and Gameful Methods for Motivation, Games for Health and Exergames

Games are recognized for their potential to be intrinsically motivating [14, 30], which has led to the rise of playful experiences and approaches like gamification [11] and serious games [1] in HCI research. These approaches implement game elements into non-game contexts and objectives outside the game content into game design to motivate and engage users [16, 23, 26]. Despite recent critical reflection [35], the theoretical foundation of research on PX is mainly based on Self-Determination Theory (SDT) [30], which states that individuals are intrinsically motivated to engage in activities that satisfy three psychological needs: competence, autonomy, and relatedness [15, 23, 30, 34]. Gameful design has shown success in motivating sustained usage of health and fitness apps and supporting therapy programs [20, 27]. Fitness games such as Dance Dance Revolution and Wii Sports, or motion-based games [32], provide a playful way for players to perform physical exercises and can motivate them to solve challenges. Fitness applications attempt to fit into individuals' daily schedules by utilizing wearable sensing technology, goal setting, and progress evaluation. To support the use of fitness games in daily life, Campbell et al. [6] suggest several design principles, including providing a challenge to the player, clear goals, and fair play. Neupane et al. [27] found that gameful designs in fitness tracker apps that motivate users to walk more address competition, challenge, social play, and real-world incentives. However, the authors noted a lack of designs that support collaboration.

## 2.2 Competitive vs. Collaborative Play and Motivation

Multiplayer games have become increasingly popular[2] and can be categorised as competitive or collaborative [38]. The Social interdependence theory suggests that competition or collaboration can affect the enjoyment of an activity and player behaviour [21]. Competitive games provide a challenge and address players' sense of competence, but they can decrease intrinsic motivation for the activity [9]. Collaborative games involve players with individual actions and a shared goal, but they are complex to design [19, 38]. Collaborative play can improve player experience and facilitate social bonds between players [10, 36]. Collaborative game mechanics has been found to be more effective in fostering motivation, adherence to programs, and personal relevance for players than competitive game elements [10, 26]. Collaborative play can also build trust between players and strengthen personal relationships [36]. Thus, designing games for health around collaborative game mechanics has the potential to yield meaningful and positive relationships with PA. However, related work has shown that even within broadly collaborative gameful design patterns, differences in acceptance and experience exist [18]. Given the sensitivities around PA in their linkage to physical abilities, appearance, and social status, this motivates investigating different modalities for communicating PA-based contributions (here: step-counts) to/as shared achievements.

## 3 IMPLEMENTATION AND METHOD

As a testbed for our research into the impact of group-based goal setting on PA, we developed the collaborative mobile game Shared Achievements. The core gameplay is built around a step-counting mechanism using the pedometer of the mobile device. In teams of 3-4 people, the players are tasked with a shared daily challenge to reach the peak of a virtual mountain. The players progress towards the peak by contributing their (physically) walked steps to the overall group goal. For the purposes of the study, two different versions of the app were used. One displayed the absolute value of steps collected by team members, while the other displayed a percentage of completion relative to the step goal of each team member. This was to help determine which method was preferable for users. To calculate the individual step goals, each player submitted their daily steps for one week prior to using the app, and the mean value plus 10 percent was then used as the individual goal. The player then contributes either the absolute number or the relative percentage of their individual goal towards their team goal. The challenges are composed of the combined individual goals of each team member. Thus, in order for the team to win a challenge, all team members should fulfill their individual goals. However, individuals are allowed to "overshoot" and compensate if other members do not reach their goals. To visualize the progress of the ongoing challenge, the main screen shows a mountain to be climbed and the accumulated progress of the team (Figure 1). To further facilitate motivation, following Chow et al. [8], the app supports communication between team members. For team communication, the app implements a simple interface for sending messages, either to the whole team or to individual members. The game was iteratively designed and evaluated by two rounds of expert feedback, and a pilot study over two weeks of play-testing. The participants came from a convenient sample, and were already acquainted. The app is developed using React Native with Expo[1] for iOS and Android. For the backend, we developed a custom Flask (Python) server with a MySQL database. The communication between the app and the backend is SSL encrypted.

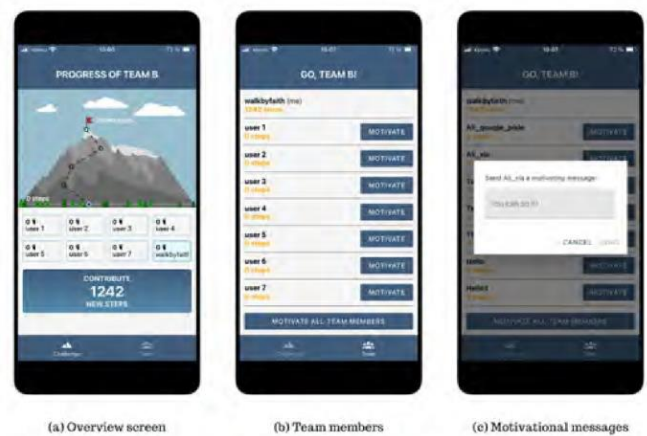

Fig. 1. App screens. a) Main screen with the team challenge and a mountain showing the team progress. Below, all team members and step contributions are shown. At the bottom is the button to contribute steps. b) Team screen depicting all members and the steps they contributed for the challenge. Next to each name is a button to send individual messages. Additionally, at the bottom is a button to message all team members. c) Dialog box to send a message.

The pre-study was conducted with 10 participants (6 female, 4 male), aged between 24 and 43 years (M=32.5). Three teams consisting of 3-4 conveniently sampled research subjects were formed. Participation lasted for two weeks: in the first week, Team 1 and Team 3 were assigned to the relative mode, whereas Team 2 was assigned to the absolute mode. In the second week, Team 2 was assigned to the relative mode, and Team 1 and Team 3 were assigned to the absolute mode. Participants were asked to collect step-counts during the week preceding the study participation and based on their daily average, the participants then set individual step goals for their daily challenge to M=4704.84 (4393.70) [min: 110, max: 16500]. The individual user challenges were completed with an average completion rate of 89.79% (87.25) [min: 0 max: 435]. The average completion rate of team goals was 69.9% (47.3) for team 1, 83.0% (51.2) for team 2, and 644.9% (2264.15) for team 3. One week after finishing the study, a post-study workshop was conducted with the study participants (8 out of 10) which was recorded and transcribed. The transcript underwent a thematic analysis in line with Braun and Clarke [5], whereby the text was coded to saturation and then grouped and iterated over with fellow team members until final themes were agreed upon and formalised.

## 4 OUTCOMES

In this section, we report on the main findings derived from the thematic analysis of the workshop transcript.

*Communication as a Key Component for Camaraderie.* Communication was found to be a key element within all interactions with the app and was an essential factor for inter-team relationships. Participants remarked that *"the most important thing for me [participant] was the group chat functionality"* as it enabled teams to work together toward the shared goal. Participants found that there was a sense of camaraderie created through the team communication, with members rallying around those who did not or could not complete the goals and asking after fellow participants. Overall, users found the message functionality in the app *"the most important thing for me [user]"*. In contrast, some participants felt that there was not enough communication between group members, which meant *"the conversation base wasn't that good"*. Participants felt it necessary but difficult to explain themselves when they were unable to complete their daily step goal, stating that *"I wanted to write to each one when I couldn't do anything, but it's too cumbersome"*. From this it was clear that the level of communication strongly influenced the team experience and the individual experience. It seems that increased communication within a team could lead to increased performance levels and greater user satisfaction.

*Surprising Intersection of Collaborative and Competitive Themes.* Participants found it more motivating to complete their goals within a group context, saying it was *"cool to compare with the team"*, and that seeing the work other people did encouraged them to complete their own goals. The users found that the experience was both *"cooperative and competitive"*, stating that it was helpful to compare with other members in the team and motivate each other through their shared achievements. Motivation within the team was a big theme, with participants finding it helpful to be completing challenges as a group and having team members to encourage and motivate them. Participants found the pressure that came from competing with team members as well as the internal pressure to improve upon their previous days' score to be positive and it kept them motivated to try and reach each day's goal. A number of participants stated that they felt guilty when they were not able to complete their goals, and *"had such a bad feeling, really"*. In small amounts this could be positive pressure, but it also appeared to negatively affect the motivation of some users. Overall, participants found the use of the application to be a good motivator in terms of increasing their PA. This came about as a result of the team dynamic, having a clear goal, and the feeling of (shared) success. Users found that it *"felt good to reach the daily goal"*, which in turn *"influenced [me] to get more steps, nice to get above the baseline"*.

*Unexpected Influence of Social Relationships and Team Dynamics.* Overall, there was a strong sense of team spirit and motivation when using the application. Users enjoyed the participatory element that was inherent in being part of a team and relationships played an important role in the overall team dynamic, with participants saying that it made a difference to be known by their team and that they *"don't want to let the team down"* as a result of the personal relationships that were formed. Participants commented on the value and importance of the group element, stating that *"the big difference is more or less about the dynamic"*. Some teams found the dynamic to be "erratic" and were sometimes "a little out of focus", which made the overall challenge more difficult. They also found that this effect was magnified when teams were smaller. A few participants remarked on the challenge of unequal contributions in the team and commented that if the environment is more competitive then *"you lose those who already have few steps"* and this may be discouraging for some participants.

*Mixed Preferences for Contribution Representations.*  Participants found the *absolute* version of the app more transparent, but also more "pushy" and confusing as *"there was only one step forward no matter how much you did"*. The difference in the baseline between team members was also seen to be very large, while also leaving progress or lack thereof more exposed. On the other hand, there was also very mixed feedback regarding the relative version of the app. Some participants found that they *"experienced more community feelings with the group"*, while others found it to be more placative and didn't like that they *"didn't know how many steps I [user] needed to take"*. Overall, there was no absolute sum difference between the two versions even though there were different preferences on an individual level.

Next to the thematic analysis, we also collected desired functionalities and features via a top-down informed analysis of the qualitative data. Users suggested a range of desired features and functionalities that they felt could enhance the experience. Many of these had to do with the goals and, for example, being able to choose different visualizations of said goal, view previous goals from history, and being able to adjust the goals themselves. One interesting point of feedback was the desire to visualize the extra work that some team members did, above and beyond their daily contribution goal. Increased social functionalities were also requested, such as ways to share more activity details with teammates (e.g. photos or location), as a way of encouraging others and engaging as a team.

## 5   DISCUSSION & CONCLUSION

The complete paper can be found with the following reference:

Young F, Alexandrovsky D, Wurhofer D, Krah, E, and Smeddinck, J. Shared Achievements: Exploring the Design of Gameful Collaborative Elements and Fostering Social Relatedness through Team Effort Contributions in a Social Physical Activity Application. Paper accepted at dHealth-2025; 2025.

## REFERENCES

[1]  Clark C. 1970. *Serious Games*. Viking Press.

[2]  Entertainment Software Association. 2017. *Essential Facts About the Computer and Video Game Industry*. Technical Report.

[3]  Max V. Birk and Regan L. Mandryk. 2018. Combating Attrition in Digital Self-Improvement Programs Using Avatar Customization. In *Proceedings of the 2018 CHI Conference on Human Factors in Computing Systems (CHI '18)*. ACM, Montreal QC, Canada, Article 660, 15 pages.  https://doi.org/10.1145/3173574.3174234

[4]  Max V. Birk, Greg Wadley, Vero Vanden Abeele, Regan Mandryk, and John Torous. 2019. Video Games for Mental Health. *Interactions* 26, 4 (June 2019), 32–36.  https://doi.org/10.1145/3328483

[5]  Virginia Braun and Victoria Clarke. 2012. Thematic Analysis. In *APA Handbook of Research Methods in Psychology, Vol 2: Research Designs: Quantitative, Qualitative, Neuropsychological, and Biological.*, Harris Cooper, Paul M. Camic, Debra L. Long, A. T. Panter, David Rindskopf, and Kenneth J. Sher (Eds.). American Psychological Association, Washington, 57–71.  https://doi.org/10.1037/13620-004

[6]  Taj Campbell, Brian Ngo, and James Fogarty. 2008. Game Design Principles in Everyday Fitness Applications. In *Proceedings of the 2008 ACM Conference on Computer Supported Cooperative Work*. ACM, San Diego CA USA, 249–252.  https://doi.org/10.1145/1460563.1460603

[7]  Colleen Cheek, Theresa Fleming, Mathijs FG Lucassen, Heather Bridgman, Karolina Stasiak, Matthew Shepherd, and Peter Orpin. 2015. Integrating Health Behavior Theory and Design Elements in Serious Games. *JMIR Mental Health* 2, 2 (April 2015), e11.  https://doi.org/10.2196/mental.4133

[8]  Clara K. Chow, Harry Klimis, Aravinda Thiagalingam, Julie Redfern, Graham S. Hillis, David Brieger, John Atherton, Ravinay Bhindi, Derek P. Chew, Nicholas Collins, Michael Andrew Fitzpatrick, Craig Juergens, Nadarajah Kangaharan, Andrew Maiorana, Michele McGrady, Rohan Poulter, Pratap Shetty, Jonathon Waites, Christian Hamilton Craig, Peter Thompson, Sandrine Stepien, Amy Von Huben, Anthony Rodgers, and on behalf of the TEXTMEDS Investigators*. 2022. Text Messages to Improve Medication Adherence and Secondary Prevention After Acute Coronary Syndrome: The TEXTMEDS Randomized Clinical Trial. *Circulation* 145, 19 (May 2022), 1443–1455.  https://doi.org/10.1161/CIRCULATIONAHA.121.056161

[9]  Edward L. Deci, Gregory Betley, James Kahle, Linda Abrams, and Joseph Porac. 1981. When Trying to Win: Competition and Intrinsic Motivation. *Personality and Social Psychology Bulletin* 7, 1 (March 1981), 79–83.  https://doi.org/10.1177/014616728171012

[10] Ansgar E. Depping and Regan L. Mandryk. 2017. Cooperation and Interdependence: How Multiplayer Games Increase Social Closeness. In *Proceedings of the Annual Symposium on Computer-Human Interaction in Play*. ACM, Amsterdam The Netherlands, 449–461. https://doi.org/10.1145/3116595.3116639

[11] Sebastian Deterding, Dan Dixon, Rilla Khaled, and Lennart Nacke. 2011. From Game Design Elements to Gamefulness: Defining "Gamification". In *Proceedings of the 15th International Academic MindTrek Conference: Envisioning Future Media Environments (MindTrek '11)*. ACM, New York, NY, USA, 9–15. https://doi.org/10.1145/2181037.2181040

[12] Theresa M. Fleming, Lynda Bavin, Karolina Stasiak, Eve Hermansson-Webb, Sally N. Merry, Colleen Cheek, Mathijs Lucassen, Ho Ming Lau, Britta Pollmuller, and Sarah Hetrick. 2017. Serious Games and Gamification for Mental Health: Current Status and Promising Directions. *Frontiers in Psychiatry* 7 (2017). https://doi.org/10.3389/fpsyt.2016.00215

[13] Viktor Gecas. 1989. The Social Psychology of Self-Efficacy. *Annual Review of Sociology* 15, 1 (Aug. 1989), 291–316. https://doi.org/10.1146/annurev.so.15.080189.001451

[14] James Paul Gee. 2007. *Good Video Games + Good Learning: Collected Essays on Video Games, Learning, and Literacy*. Peter Lang.

[15] Stuart Hallifax, Audrey Serna, Jean-Charles Marty, Guillaume Lavoué, and Elise Lavoué. 2019. Factors to Consider for Tailored Gamification. In *Proceedings of the Annual Symposium on Computer-Human Interaction in Play*. ACM, Barcelona Spain, 559–572. https://doi.org/10.1145/3311350.3347167

[16] Juho Hamari, Jonna Koivisto, and Harri Sarsa. 2014. Does Gamification Work? – A Literature Review of Empirical Studies on Gamification. In *2014 47th Hawaii International Conference on System Sciences*. IEEE, Waikoloa, HI, 3025–3034. https://doi.org/10/gfj8gf

[17] William L. Haskell, I-Min Lee, Russell R. Pate, Kenneth E. Powell, Steven N. Blair, Barry A. Franklin, Caroline A. Macera, Gregory W. Heath, Paul D. Thompson, and Adrian Bauman. 2007. Physical Activity and Public Health: Updated Recommendation for Adults from the American College of Sports Medicine and the American Heart Association. *Medicine & Science in Sports & Exercise* 39, 8 (Aug. 2007), 1423–1434. https://doi.org/10.1249/mss.0b013e3180616b27

[18] Robert Hermann, Marc Herrlich, Dirk Wenig, Jan Smeddinck, and Rainer Malaka. 2013. Strong and Loose Cooperation in Exergames for Older Adults with Parkinson's Disease. *Mensch & Computer 2013-Workshopband* (2013). https://doi.org/10.1524/9783486781236.249

[19] Carolina Islas Sedano, Maira B. Carvalho, Nicola Secco, and C. Shaun Longstreet. 2013. Collaborative and Cooperative Games: Facts and Assumptions. In *2013 International Conference on Collaboration Technologies and Systems (CTS)*. IEEE, San Diego, CA, 370–376. https://doi.org/10.1109/CTS.2013.6567257

[20] Daniel Johnson, Sebastian Deterding, Kerri-Ann Kuhn, Aleksandra Staneva, Stoyan Stoyanov, and Leanne Hides. 2016. Gamification for Health and Wellbeing: A Systematic Review of the Literature. *Internet Interventions* 6 (Nov. 2016), 89–106. https://doi.org/10.1016/j.invent.2016.10.002

[21] David W Johnson and Roger T Johnson. 1989. *Cooperation and Competition: Theory and Research*. Interaction Book Company.

[22] Robert B. Kelly, Stephen J. Zyzanski, and Sonia A. Alemagno. 1991. Prediction of Motivation and Behavior Change Following Health Promotion: Role of Health Beliefs, Social Support, and Self-Efficacy. *Social Science & Medicine* 32, 3 (Jan. 1991), 311–320. https://doi.org/10.1016/0277-9536(91)90109-P

[23] Jonna Koivisto and Juho Hamari. 2019. The Rise of Motivational Information Systems: A Review of Gamification Research. *International Journal of Information Management* 45 (April 2019), 191–210. https://doi.org/10/gf9ck9

[24] Andrés Lucero, Evangelos Karapanos, Juha Arrasvuori, and Hannu Korhonen. 2014. Playful or Gameful?: creating delightful user experiences. *interactions* 21 (05 2014), 34–39. https://doi.org/10.1145/2590973

[25] Elisa Mekler, Florian Brühlmann, Alexandre Tuch, and Klaus Opwis. 2017. Towards understanding the effects of individual gamification elements on intrinsic motivation and performance. *Computers in Human Behavior* 71 (06 2017). https://doi.org/10.1016/j.chb.2015.08.048

[26] Benedikt Morschheuser, Juho Hamari, and Alexander Maedche. 2019. Cooperation or Competition – When Do People Contribute More? A Field Experiment on Gamification of Crowdsourcing. *International Journal of Human-Computer Studies* 127 (July 2019), 7–24. https://doi.org/10.1016/j.ijhcs.2018.10.001

[27] Aatish Neupane, Derek Hansen, Jerry Alan Fails, and Anud Sharma. 2021. The Role of Steps and Game Elements in Gamified Fitness Tracker Apps: A Systematic Review. *Multimodal Technologies and Interaction* 5, 2 (Jan. 2021), 5. https://doi.org/10.3390/mti5020005

[28] Olga Perski, Ann Blandford, Robert West, and Susan Michie. 2017. Conceptualising Engagement with Digital Behaviour Change Interventions: A Systematic Review Using Principles from Critical Interpretive Synthesis. *Translational Behavioral Medicine* 7, 2 (June 2017), 254–267. https://doi.org/10.1007/s13142-016-0453-1

[29] Gregory N. Ruegsegger and Frank W. Booth. 2018. Health Benefits of Exercise. *Cold Spring Harbor Perspectives in Medicine* 8, 7 (July 2018), a029694. https://doi.org/10.1101/cshperspect.a029694

[30] Richard M Ryan and Edward L Deci. 2000. Self-Determination Theory and the Facilitation of Intrinsic Motivation, Social Development, and Well-Being. *American Psychologist* (2000), 67. https://doi.org/10.1037/0003-066x.55.1.68

[31] Jan D. Smeddinck. 2016. Games for Health. In *Entertainment Computing and Serious Games*, Ralf Dörner, Stefan Göbel, Michael Kickmeier-Rust, Maic Masuch, and Katharina Zweig (Eds.). Lecture Notes in Computer Science, Vol. 9970. Springer International Publishing, Cham, 212–264. http://link.springer.com/10.1007/978-3-319-46152-6_10 "projects":["adaptify","sdf"],"url_preprint": "./files/papers/b03_gamesForHealth_SELF_ARCH_IV_En_.pdf ".

[32] Jan D. Smeddinck. 2016. Games for Health. In *Entertainment Computing and Serious Games*, Ralf Dörner, Stefan Göbel, Michael Kickmeier-Rust, Maic Masuch, and Katharina Zweig (Eds.). Vol. 9970. Springer International Publishing, Cham, 212–264. https://doi.org/10.1007/978-3-319-46152-6_10

[33] John M. Tauer and Judith M. Harackiewicz. 2004. The Effects of Cooperation and Competition on Intrinsic Motivation and Performance. *Journal of Personality and Social Psychology* 86, 6 (2004), 849–861. https://doi.org/10.1037/0022-3514.86.6.849

[34] April Tyack and Elisa D. Mekler. 2020. Self-Determination Theory in HCI Games Research: Current Uses and Open Questions. In *Proceedings of the 2020 CHI Conference on Human Factors in Computing Systems*. ACM, Honolulu HI USA, 1–22. https://doi.org/10.1145/3313831.3376723

[35] April Tyack and Elisa D. Mekler. 2024. Self-Determination Theory and HCI Games Research: Unfulfilled Promises and Unquestioned Paradigms. arXiv:2405.12639 [cs]

[36] Kellie Vella, Daniel Johnson, and Leanne Hides. 2015. Playing Alone, Playing With Others: Differences in Player Experience and Indicators of Wellbeing. In *Proceedings of the 2015 Annual Symposium on Computer-Human Interaction in Play*. ACM, London United Kingdom, 3–12. https://doi.org/10.1145/2793107.2793118

[37] WHO. 2020. *WHO Guidelines on Physical Activity and Sedentary Behaviour*. World Health Organization, Geneva.

[38] José P. Zagal, Jochen Rick, and Idris Hsi. 2006. Collaborative Games: Lessons Learned from Board Games. *Simulation & Gaming* 37, 1 (March 2006), 24–40. https://doi.org/10.1177/1046878105282279